\documentclass[twocolumn,showpacs,preprintnumbers,amsmath,amssymb]{revtex4}

\usepackage{graphicx}
\usepackage{dcolumn}
\usepackage{bm}

\begin{document}
\title{Drastic change in transport of entropy with quadrupolar ordering in PrFe$_{4}$P$_{12}$  }
\author{A. Pourret$^{1}$, K. Behnia$^{1}$, D.
Kikuchi$^{2}$, Y. Aoki$^{2}$, H. Sugawara$^{3}$  and H.
Sato$^{2}$} \affiliation{(1)Laboratoire de Physique
Quantique(CNRS), ESPCI, 10 Rue de Vauquelin,
75231 Paris, France \\
(2)Department of Physics, Tokyo Metropolitan University, Tokyo
192-0397, Japan \\
(3)Faculty of Integrated Arts and Science, University of
Tokushima, Tokushima 770-8502, Japan}

\date{February 23, 2006}

\begin{abstract}
The antiferroquadrupolar ordering of PrFe$_{4}$P$_{12}$ is
explored by probing thermal and thermoelectric transport. The
lattice thermal conductivity drastically increases with the
ordering, as a consequence of a large drop in carrier
concentration  and a strong electron-phonon coupling. The low
level of carrier density in the ordered state is confirmed by the
anomalously large values of the Seebeck and Nernst coefficients.
The results are reminiscent of URu$_{2}$Si$_{2}$ and suggest that
both belong to the same class of partial metal-insulator
transitions. The magnitude of the Nernst coefficient, larger than
in any other metal, indicates a new route for Ettingshausen
cooling at Kelvin temperatures.
\end{abstract}


\maketitle

\begin{figure}
{\includegraphics{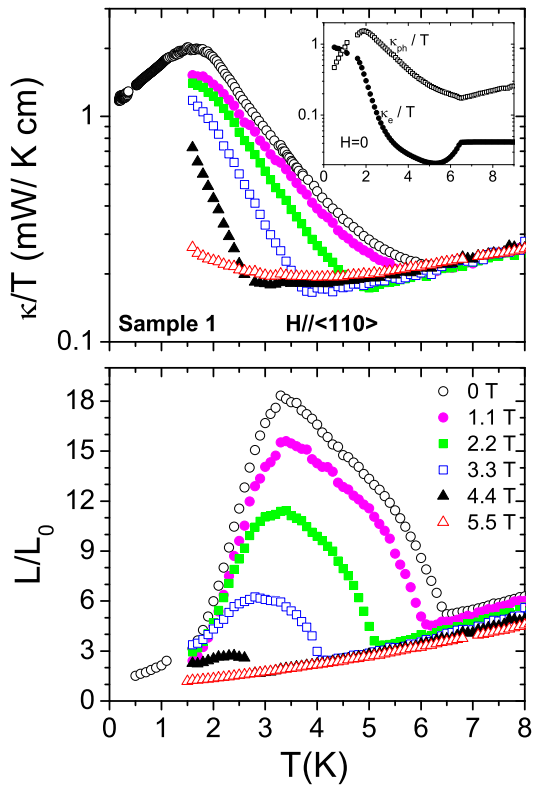}} \caption{\label{fig1}Temperature
dependence of the thermal conductivity divided by temperature
(upper panel) and the normalized Lorenz number(lower panel) for
different magnetic fields in PrFe$_{4}$P$_{12}$. The inset in
upper panel shows the temperature dependence of the phononic and
electronic contributions to thermal conductivity at zero field
assuming the validity of the Wiedemann-Franz law in the whole
temperature range.}
\end{figure}

In Pr intermetallics, the Kondo coupling between $f$-electrons and
conduction electrons, the core of heavy-fermion phenomena,
presents unique features associated with the double occupancy of
the $4f$ orbital. In particular, the filled Skutterudites of the
PrTr$_{4}$Pn$_{12}$ family (where Tr stands for Fe, Ru or Os and
Pn stands for P, As or Sb) have become a growing subject of
investigation\cite{aoki1}. The diversity of electronic
instabilities associated with this particular structure (including
superconducting\cite{bauer}, insulating\cite{seikine}and
metallic\cite{sato}) is intriguing. The extreme sensitivity of the
ground state to small variations suggests a subtle interplay
between the orbital degrees of freedom, the crystal electric field
and the incomplete hybridization between $f$ electrons and the
conduction band\cite{otsuki}.

In this context, the case of PrFe$_{4}$P$_{12}$ deserves
particular attention. This system is host to a phase transition at
6.5 K with sharp signatures in all macroscopic
properties\cite{sato,aoki2,sugawara}. Contrary to what was
initially assumed\cite{torikachvili}, it does not correspond to an
antiferromagnetic (AFM) ordering. In the ordered state, no
magnetic Bragg peak is detected by neutron
scattering\cite{keller,hao} and no internal field is seen by muon
spin relaxation measurements\cite{aoki1}. Moreover, nuclear
specific heat data rule out any ordered magnetic moment on the Pr
site\cite{aoki2}. All these experiments converge to establish the
non-magnetic nature of this phase transition. An alternative order
parameter, namely an antiferroquadrupolar(AFQ) one has emerged as
the most plausible candidate for the ordered state\cite{aoki1}.
This scenario would explain\cite{curnoe1} both the field-induced
staggered magnetic moment\cite{hao} and the superlattice
reflection resolved by X-ray diffraction\cite{iwasa}. Both of them
correspond to the same wave-vector q$_{A}$=[1,0,0] which,
according to the band calculations\cite{harima}, is a possible
nesting vector of the Fermi surface. The AFQ scenario is also
compatible with the magnetization data\cite{aoki2,kiss}.

In this Letter, we report on a study of thermal and thermoelectric
transport in PrFe$_{4}$P$_{12}$, which documents a decimation of
the Fermi surface with the ordering. According to our analysis,
most carriers disappear at the transition, paving the way for an
enhanced mean-free-path of both phonons and the residual
quasi-particles. Thus, the ordered state of PrFe$_{4}$P$_{12}$
emerges as a heavy-fermion semi-metal. Such a radical interplay
between itinerant electrons and the loss of orbital degrees of
freedom contrasts with typical cases of AFQ ordering, such as
CeB$_{6}$ or DyB$_{2}$C$_{2}$. It is reminiscent of what was
observed in the case of the hidden-order transition in
URu$_{2}$Si$_{2}$\cite{bel1,behnia1,sharma} and provides useful
input for the ongoing debate on the identity of hidden order
there.

\begin{figure}
{\includegraphics{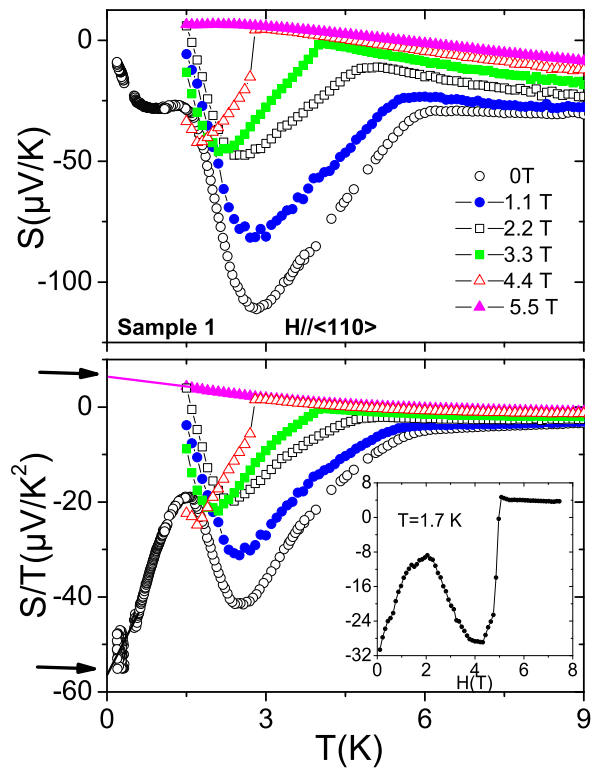}} \caption{\label{fig2}Temperature
dependence of $S$, the Seebeck coefficient(upper panel),  and
$S/T$(lower panel) for different magnetic fields in
PrFe$_{4}$P$_{12}$. Solid lines are linear extrapolation of the
data used to extract the T=0 value of $S/T$ as indicated by the
arrows. The inset shows the field dependence of $S/T$.}
\end{figure}
Single crystals of PrFe$_{4}$P$_{12}$ were grown by a tin-flux
method described elsewhere\cite{torikachvili,sugawara}.  Nernst
effect, thermopower and thermal conductivity were measured using a
one-heater-two-thermometer set-up which allowed us to measure all
transport coefficients of the sample in the same conditions.  For
temperatures above 1.5 K  a set-up with cernox thermometers in a
$^{4}$He cryostat was used. The data were complemented with
subkelvin measurements using another set-up with RuO$_{2}$
thermometers in a dilution refrigerator. The magnetic field,
$\mathbf{H}$ was applied perpendicular to the [applied]
heat-current ($\mathbf{J_{Q}}$) and the [measured] electric-field
($\mathbf{E}$) vectors. For sample no.1 (which had a residual
resistivity of $\rho_{0}= 23\mu \Omega$cm), the configuration was
$\mathbf{H}\parallel [1,1,0] ; \mathbf{J_{Q}}\parallel
\mathbf{E_{x}}\parallel[0,0,1]$ and
$\mathbf{E_{y}}\parallel[-1,1,0]$. For sample no.2 ($\rho_{0}=
32\mu \Omega$cm), it was $\mathbf{H}\parallel[0,0,1];
\mathbf{J_{Q}}\parallel \mathbf{E_{x}}\parallel[1,-1,0]$ and
$\mathbf{E_{y}}\parallel[1,1,0]$. Hence, Nernst signal ($N =
\frac{E_{y}}{\nabla_{x}T}$) and Seebeck coefficient ($S
=\frac{E_{x}}{\nabla_{x}T}$) were measured in adiabatic
conditions.

Fig. 1 presents the results of thermal conductivity($\kappa$)
measurements. As seen in the upper panel, the onset of ordering is
marked by a steep increase in $\kappa/T$. This enhancement is
diminished by the application of a magnetic field which leads to a
gradual suppression of the ordering. The effect of ordering on
lattice and electronic thermal conductivities can be probed by
checking the temperature dependence of the Lorenz number
(L=$\frac{\kappa \rho}{T}$). The lower panel presents the
temperature dependence of L/L$_{0}$ (where L$_{0}$=2.44$\times
10^{-8}$ V$^{2}$/K$^{2}$). Its magnitude at T$_{Q}$ ($\sim 5$ )
quantifies the dominance of the lattice contribution to heat
transport. The transition is accompanied by a drastic enhancement
in L/L$_{0}$ which reaches a peak value of 18 and then decreases
to yield unity in the zero-temperature limit, in agreement with
the Wiedemann-Franz (WF) law. If electrons were the only heat
carriers, then L/L$_{0} <1 $ at finite temperatures and its
magnitude would decrease with magnetic ordering. Indeed, the
presence of magnetic fluctuations above T$_{N}$ tends to amplify
the relative weight of large-\textbf{q} scattering and to rectify
the excess in thermal resistivity produced by the presence of
small-\textbf{q} inelastic scattering\cite{paglione}. The unusual
enhancement of L/L$_{0}$ observed here is reminiscent of (but more
dramatic than) what was observed in
URu$_{2}$Si$_{2}$\cite{behnia1} and could be safely attributed to
an increase in the \emph{lattice} heat transport. We put aside the
unlikely hypothesis of a supplementary heat transport by exotic
excitations in the ordered state. Indeed, in sharp contrast to
what is reported here, heat transport in CeB$_{6}$ is not affected
by AFQ ordering \cite{marcenat}.

The relative weights of electronic, $\kappa_{e}$ and phononic,
$\kappa_{ph}$ contributions to heat transport can be estimated
assuming that $\kappa_{e}$ follows the WF law even at finite
temperatures. Since $\frac{\kappa_{e}\rho}{T}$ can become as low
as $0.4 L_{0}$\cite{paglione} in presence of inelastic scattering,
this procedure would only indicate the qualitative trend. As seen
in the inset of the upper panel, both components of heat transport
are affected by the ordering. While $\kappa_{ph}/T$ displays a
steady enhancement of one order of magnitude, $\kappa_{e}/T$
increases significantly after an initial decrease. This can be
understood in a picture similar to the one proposed for
URu$_{2}$Si$_{2}$\cite{behnia1}. The opening of a gap at the onset
of ordering destroys most of the Fermi surface. The associated
drop in the density of itinerant electrons leads to a decrease in
the phonon scattering and an enhancement of the lattice thermal
conductivity. The behavior of $\kappa_{e}/T$ reflects the
combination of a decrease, induced by the drop in the carrier
number, and an increase due to the rise in the mean-free-path of
the surviving quasi-particles as a consequence of the restricted
phase space in the ordered state. Thus, the analysis of thermal
conductivity underlines the strength of the electron-phonon
coupling in PrFe$_{4}$P$_{12}$ and points to a large decrease in
the carrier density with ordering.
\begin{figure}
{\includegraphics{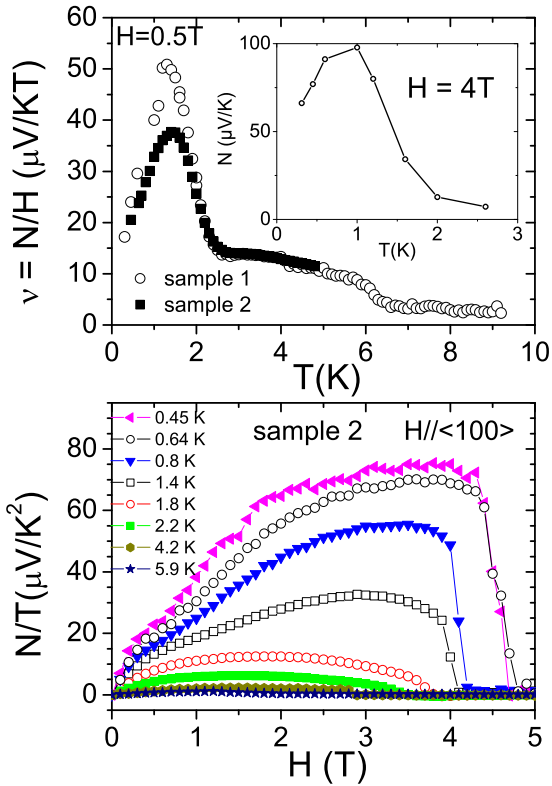}} \caption{\label{fig3}Upper panel:
Temperature dependence of the Nernst coefficient, $\nu= N/H$ for
two different samples at 0.5 T. The inset shows the temperature
dependence of $N$ for sample no. 1 at 4 T. Lower panel: The field
dependence of the Nernst signal divided by temperature, $N/T$, for
different magnetic fields in sample 2.}
\end{figure}
Fig. 2 presents the data on the Seebeck coefficient. In agreement
with the earliest study\cite{sato}, the absolute value of
thermopower, $S$, was found to increase in the ordered state and
attain a large peak (-113 $\mu$V/K).  By continuing the zero-field
measurements down to 0.18 K,  a shoulder-like feature around 1 K
in $S(T)$ is also resolved. The application of the magnetic field
leads to a gradual suppression of the enhancement in $S$. The
lower panel, which presents the same data as $S/T$ \emph{vs.}
temperature, reveals a non-monotonous $S/T$ at zero field which
extrapolates to $-56\pm 8\mu$V/K$^{2}$ at zero temperature. Its
absolute value is comparable to the largest $S/T$ ever reported
($+50 \mu$V/K$^{2}$ in CeNiSn\cite{hiess}). In many correlated
metals, the absolute value of the dimensionless ratio $q =
\frac{S}{T} \frac{N_{Av} e}{\gamma}$ ($\gamma$ is the electronic
specific heat, $e$ the elementary charge and $N_{Av}$ the Avogadro
number) is of the order of unity\cite{behnia2}. Miyake and Kohno
have argued that, in the zero-temperature limit, for both Born and
unitary limits of scattering, $S/T$ becomes inversely proportional
to the renormalized Fermi Energy and this leads to the observed
correlation\cite{miyake}. Let us recall that when the carrier
density is much lower than one itinerant electron per formula
unit, a proportionally larger $|q|$ is expected\cite{behnia2}. For
example, in the case of CeNiSn, $q\simeq107$ is compatible with
the very low level of carrier density ($< 0.01 e^{-}$/f.u.) in
this heavy-fermion semi-metal. The relative magnitudes of $S/T$
and $\gamma$ is compatible with the low level of carrier density
in the ordered state of PrFe$_{4}$P$_{12}$. At zero-field, $S/T$
and $\gamma\simeq$0.1 J/(K$^{2}$mol)\cite{aoki2} yield
$q\simeq-58\pm10$. At B = 5.5 T, by linearly extrapolating the
low-temperature data to T=0, one obtains $S/T \simeq
+8\pm2\mu$V/K$^{2}$, and this together with $\gamma$ (estimated to
be $\simeq$1.3 J/(K$^{2}$mol) for this orientation\cite{namiki}),
yield $q\simeq 0.6\pm0.2$. This simple argument indicates that the
high-field state is an ordinary HF metal, and the ordered state is
a dilute (with $\sim$0.02 carriers per f.u.) liquid of heavy
quasi-particles, i.e. a heavy-fermion semi-metal.

Fig. 3 displays the temperature dependence of the Nernst
coefficient, $\nu$, which becomes very large in the ordered state,
in particular below T$_{x}\sim$ 2.8 K and reaches a maximum,
$\nu_{max}$, around $T\sim$ 1.5 K, which is $51\pm3\mu$V/KT in
sample 1 and $38\pm2\mu$V/KT in sample 2. Since
$\nu^{1}_{max}/\nu^{2}_{max}\sim\rho^{2}_{0}/\rho^{1}_{0}\sim1.3$,
$\nu_{max}$ appears to inversely scale with $\rho_{0}$. These
numbers are one order of magnitude larger than what was found in
URu$_{2}$Si$_{2}$\cite{bel1} or in CeCoIn$_{5}$\cite{bel2}, which
are host to Nernst coefficients of exceptionally large magnitudes.
The lower panel of the same figure presents the field dependence
of the Nernst signal and shows that the destruction of the order
by the magnetic field leads to the suppression of $N$ down to
negligibly low values. The giant Nernst signal is indeed a
property of the ordered state of PrFe$_{4}$P$_{12}$.

How can quasi-particles of a \emph{non-magnetic} metal produce a
Nernst signal of such a magnitude? This question brings us to the
following relation in the Boltzmann
picture\cite{wang,oganesyan,bel1}:
\begin{equation}\label{1}
N= \frac{\pi^{2}}{3}\frac{k_{B}^{2}T}{e}\frac{\partial
\\tan\theta_{H}}{\partial \epsilon}|_{ \epsilon_{F}}
\end{equation}

In other words, the Nernst signal tracks the energy dependence of
the tangent of the Hall angle
($\tan\theta_{H}=\frac{\rho_{xy}}{\rho_{xx}}$) at the Fermi level.
Since in a first approximation, this energy derivative is just
proportional to $\frac{\tan\theta_{H}}{\epsilon_{F}}$, a
combination of a large Hall angle and a small Fermi energy can
lead to a very large Nernst signal. Comparing the temperature
dependence of $N/T$ and $\tan\theta_{H}$ in Fig. 4 reveals the
relevance of this picture to PrFe$_{4}$P$_{12}$. The two
quantities display a striking correlation. Ordering is associated
with a jump in $\tan\theta_{H}$ which increases by two orders of
magnitude and a (much smaller) increase in $N/T$. Another distinct
increase in both quantities is also visible at T$_{x}\sim$ 2.8 K.
The unprecedented magnitude of the Nernst coefficient in the
ordered state of PrFe$_{4}$P$_{12}$ is thus concomitant with a
large value of $\tan\theta_{H}$. Quantitatively, according to the
approximation derived from eq. 1, the low-temperature magnitudes
of $N/T$ and $\tan\theta_{H}$ yield $\epsilon_{F}\sim$7 K,
comparable to the width of the Kondo resonance (8.7 K) estimated
by specific heat\cite{aoki2}.
\begin{figure}
\resizebox{!}{0.25\textwidth}{\includegraphics{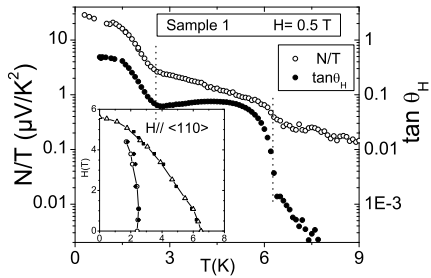}}
\caption{\label{fig4} Comparison of the temperature dependence of
the Nernst signal divided by temperature and the tangent of the
Hall angle for a field along $<110>$. The two vertical dot lines
mark the two temperature scales, T$_{Q}$ and T$_{x}$. The inset
shows the phase diagram for the same field orientation. The
boundary of the ordered state determined by
magnetization(\cite{aoki1}) and thermal conductivity
($\vartriangle$) is shown. A second line is detected by additional
anomalies in Seebeck($\circ$) and Nernst ($\blacklozenge$)
coefficients.}
\end{figure}

The phase transition in PrFe$_{4}$P$_{12}$ is a case of partial or
\emph{aborted} metal-insulator transition (MIT), where the Fermi
surface(FS) is not completely wiped out\cite{harima2}.  Band
calculations\cite{harima}have found that the main FS is a
distorted cube with a strong nesting instability. Experimentally,
de Haas van Alphen measuements\cite{sugawara} detected in the
ordered state (i.e. for $H < 5 T$) a low-frequency branch
corresponding to a FS occupying only 0.15 of the first Brillouin
zone. This would yield a carrier density of 0.003 holes/f.u. which
is compatible with the magnitude of the Hall coefficient at low
temperatures (yielding 0.005 holes/f.u.), but significantly lower
than the carrier density (0.02 $e^{-}$/f.u.) estimated from the
magnitude of $q$. This discrepancy points to the presence of other
undetected pockets of FS in the ordered state as initially
suggested by the contrast between the large value of $\gamma$ and
the modest mass enhancement of the only FS detected by
dHvA\cite{sugawara}. The unobserved electron-like quasi-particles
should be heavier with a shorter mean-free-path. These two
assumptions are sufficient to explain why they have escaped
detection by dHvA and why they dominate specific heat and
thermopower but contribute little to the Hall effect. Assuming
that the latter is entirely due to a single FS, one may write :
$\tan\theta_{H}=\frac{1}{\ell_{B}^{2}}\frac{\ell_{e}}{k_{F}}$
(where $k_{F}$ and $\ell_{e}$ are respectively the Fermi
wave-vector and the electronic mean-free-path and
$\ell_{B}=\sqrt{\frac{e}{\hbar B}}$ is the magnetic length scale).
The enhancement of $\tan\theta_{H}$ at T$_{Q}$ reflects combined
effect of the shrinking of the Fermi surface and the increase in
the mean-free-path. Taking for $k_F$ the average radius of the
single FS seen by dHvA measurements\cite{sugawara}, the magnitude
of $\tan\theta_{H}$ at low temperatures implies $\ell_{e}\sim
4000\AA$, compatible with the relatively easy observation of
quantum oscillations at 3 T\cite{sugawara}. Very recently, a
complete MIT induced by pressure was observed in
PrFe$_{4}$P$_{12}$\cite{hidaka}. Remarkably, both the
pressure-induced insulator and the ambient pressure AFQ are
destroyed in presence of a magnetic field of comparable magnitude.
However, X-ray diffraction measurements under
pressure\cite{kawana} resolve a superlattice reflection present in
the AFQ state and none in the pressure-induced insulator pointing
to a different mechanism for the two transitions.

Both $N/T$ and $\tan\theta_{H}$ show a steep increase below
T$_{x}$($\sim$ 2.8 K). This temperature is marked by a minimum in
$S/T$ (see Fig. 2), a broad shoulder-like feature in specific
heat\cite{matsuda,aoki2} and a minimum in bulk
magnetization\cite{aoki2} close to this temperature. This energy
scale was observed for both field orientations studied here and
does not vary much with the magnetic field (see the inset in
Fig.4). It is too early to identify it as another phase transition
in the ordered state since the feature in specific heat is just a
broad shoulder.

A number of implications emerge. The  consequences of ordering on
transport properties in PrFe$_{4}$P$_{12}$ and in
URu$_{2}$Si$_{2}$ are similar. In both cases, the transition leads
to a drastic, yet incomplete, destruction of the FS\cite{tayama}.
Moreover, according to our analysis, the exceptional magnitude of
the Nernst coefficient is just a consequence of three independent
factors: a low carrier density, a large mass enhancement and a
long mean-free-path. Experiments on heavy-fermion semi-metals
would tell if these are the only ingredients needed to produce a
Nernst signal of such a magnitude, or there is any additional
source specific to the exotic orders. An Ettingshausen
cooler\cite{scholz} is conceivable when the thermomagnetic figure
of merit ($Z'T= \frac{N^{2}}{L}$) is close to unity. For a metal,
this requirement means $N \approx \sqrt{L_{0}}= 153 \mu$V/K. To
the best of our knowledge, PrFe$_{4}$P$_{12}$ is the first case of
a metal approaching this value at 1 K.

We thank D. Jaccard and J. Flouquet for helpful comments. This
work was supported by the French ICENET project
(ANR-05-BLAN-0055-02) and by a Grant-in-Aid for Scientific
Research Priority Area "Skutterudite" (No. 15072206) of MEXT,
Japan.

\end{document}